# Using CAS to Manage Role-Based VO Sub-Groups

S. Cannon, S. Chan, D. Olson, C. Tull
*Lawrence Berkeley National Laboratory, Berkeley, CA*

V. Welch
*Argonne National Laboratory, Argonne, IL*

L. Pearlman
*USC Information Sciences Institute, Marina del Rey, CA*

We have developed and tested a prototype VO-Role management system using the Community Authorization Service (CAS) from the Globus project. CAS allows for a flexible definition of resources. In this prototype we define a role as a resource within the CAS database and assign individuals in the VO access to that resource to indicate their ability to assert the role. The access of an individual to this VO-Role resource is then an annotation of the user's CAS proxy certificate. This annotation is then used by the local resource managers to authorize access to local compute and storage resources at a granularity that is based on neither VOs nor individuals. We report here on the configuration details for the CAS database and the Globus Gatekeeper and on how this general approch could be formalized and extended to meet the clear needs of LHC experiments using the Grid.

## 1. INTRODUCTION

LHC-era HENP experiments will generate unprecedented volumes of data and require commensurately large compute resources. These resources are larger than can be marshaled at any one site within the community. Production reconstruction, analysis, and simulation will need to take maximum advantage of these distributed computing and storage resources using the new capabilities offered by the Grid computing paradigm. Since large-scale, coordinated Grid computing involves user access across many Regional Centers and national and funding boundaries, one of the most crucial aspects of Grid computing is that of user authentication and authorization. While projects such as the DOE Grids CA have gone a long way to solving the problem of distributed authentication, the authorization problem is still largely open.

One model of authorization is the presentation of a group membership credential by the user to the resource provider. The resource provider then translates that group membership into a set of local rights to be granted to that user. In the case of Grid computing, these groups will often correspond to the user's membership in a Virtual Organization (VO). In HENP the Grid VOs logically map onto entire experiments. I.E. The four LHC experiments (ALICE, ATLAS, CMS, LHCb) would map onto four separate VOs. However, a finer granularity of membership is needed to define sub-groups within a VO which map onto groups and/or individuals performing specific roles within the respective experiments. This VO-Role membership attribute is not tied to a particular individual nor applied to the Virtual Organization as a whole.

We have developed and tested a prototype VO-Role management system using the Community Authorization Service (CAS) from the Globus project. CAS allows for a flexible definition of resources. In this prototype we define a role as a resource within the CAS database and assign individuals in the VO access to that resource to indicate their ability to assert the role. The access of an individual to this VO-Role resource is then an annotation of the user's CAS proxy certificate. This annotation is then used by the local resource managers to authorize access to local compute and storage resources at a granularity that is based on neither VOs nor individuals. We report here on the configuration details for the CAS database and the Globus Gatekeeper and on how this general approch could be formalized and extended to meet the clear needs of LHC experiments using the Grid.

## 2. OVERVIEW OF CAS

The Community Authorization Service (CAS) [8][9] is a system developed by the Globus Project to allow virtual organizations (VOs) to flexibly and expressively authorize access to resources and data in large distributed Grids. Since the introduction of CAS in March of 2002, CAS has undergone significant changes based on requirements feedback from a number of HEP resource sites.

The CAS architecture builds on public key authentication and delegation mechanisms provided by the Globus Toolkit [7] Grid Security Infrastructure (GSI) [2][6], a widely used set of authentication and authorization mechanisms that address single sign on, delegation, and credential mapping issues that arise in VO settings.

The Grid Security Infrastructure (GSI) software is a set of libraries and tools that allow users and applications to access resources securely. GSI focuses primarily on authentication and message protection, defining single sign-on algorithms and protocols, cross-domain authentication protocols, and delegation mechanisms for creating temporary credentials for users and for processes executing on a user's behalf. GSI is based on Public Key Infrastructure (PKI) and uses authentication credentials composed of X.509 [3] certificates and private keys. In brief, a GSI user generates a public-private key pair and obtains an X.509 certificate from a trusted entity called a Certificate Authority (CA). These credentials then form the basis for authenticating the user to resources on the Grid.

GSI now uses temporary credentials called proxy credentials. Proxy credentials allow GSI to support single sign-on by allowing users to access resources at multiple sites without repeated authentication and to delegate their rights to remote processes. This single sign-on capability is critical to advanced Grid applications in which a single interaction may involve the coordinated use of resources at many locations.

## 3. RIGHTS GRANULARITY

Authorization and authentication are critically important issues in realizing the vision of Grid computing. However, with few exceptions, discussions of philosophy and implementation for authorization and authentication of users on the Grid have focused exclusively on two levels of rights granularity. These two levels of rights granularity are at the extremes of the A&A domain, ie. rights granted to the individual and rights granted to the VO as a whole.

Authentication typically involves verifying an individual's identity and her/his membership in a VO. These two extremes (the individual and the VO) can be adequate for relatively small VOs. But in a typical LHC era experiment where the VO is comprised of up to 2000 individuals, an intermediate granularity of access control is required.

We distinguish two intermediate granularities of access rights which we call roles and groups. HENP experiments typically contain sub-groupings of collaborators which we call groups. These groups can be physics groups studying a particular physics channel or physics process, or they can be sub-detector groups with special responsibility for a particular detector subsystem of the overall experimental apparatus (eg. In ATLAS the Liquid Argon Calorimeter or the Muon Chambers). These groups share information, code, and files at a higher rate than the rate found across such group boundaries. As well, they often have a collective group responsibility (eg. Producing and maintaining detector calibration databases.) which can map onto specific access rights to distributed Grid resources.

HENP experiments also contain specific roles which are adopted for a limited time by individuals or small teams within the collaboration. Examples of such roles include the Software Librarian responsible for maintaining the central software repository of the experiment, the Release Coordinator responsible for defining, coordinating, and building the stable experimental releases of software, and the Production Manager responsible for running and managing the production runs of experimental data reduction and analysis.

These roles are typically assigned to an individual within the collaboration for a limited term, and then reassigned to another individual as a way of spreading both responsibility and experience across the collaboration. These roles come with specific responsibilities and access rights for a wide variety of compute resources (such as files, databases, share/priorities of compute and storage allocation usage, etc).

The distinction between groups and roles is that roles typically map onto one or very few individuals but change frequently, while groups map onto a larger subsection of the collaboration (dozens or even hundreds of individuals) but remain relatively stable (Members join a group at some rate, but the overall population remains otherwise effectively static.).

## 4. PROTOTYPE PROJECT

### 4.1. Goal of Project

CAS is being developed partly under the auspices of the Particle Physics Data Grid (PPDG) project. Although CAS was not specifically designed and developed to address the issue of sub-VO role-based access control, it was explicitly developed to be flexible and customizable to meet all aspects of a VO's A&A needs in a Grid environment.

The goal of this project was to take the authorization mechanism being developed within PPDG and demonstrate that it is capable of managing sub-VO roles and groups without fundamentally changing the mechanism nor the associated tools.

To demonstrate that CAS can handle the requirements of sub-VO roles, we needed to demonstrate two things:
1> The granting and rescinding of the rights associated with that role to one, and then a different user.
2> The use of the rights associated with the role to access role-specific resources, and the denial of access to those resources when the rights have been rescinded.

To demonstrate that CAS can handle the requirements of sub-VO groups, we needed to demonstrate granting rights associated with that group to multiple users (ie. members of the group), and that the these multiple users can then use these rights to access resources equally.

For simplicity, we chose to grant access to file resources via GSIFTP for our tests.

### 4.2. CAS Credential Changes

#### 4.2.1. Credential Contents

Detailed fully in [8] and [9], the purpose of a CAS server is to authenticate users as a member of a VO and issue to them a cryptographically signed assertion that allows the user to assert to third parties that the user is a VO member and what their rights should be in regards to VO policy.

In the current CAS prototype [3], these rights are specific access rights on files or directory trees - e.g. the user can read file /tmp/foo, can write to /tmp/bar/*, etc. - however the CAS policy is flexible in this regard and is capable of expressing any right that can be expressed as a <action, target> tuple. By configuration of the CAS database new sets of actions and types of targets can be easily added.

For our prototype we defined a new set of actions and targets to allow expressing role membership in the tuples. A new action was defined to indicate role membership and a new target type was defined to identify specific roles. In our prototype we defined roles with a two-level hierarchy to allow for the expression of both the VO that the role was associated with and the role's name - for example "atlas/admin" to indicate an administrator in the Atlas VO. While we experimented with only two levels in the role names, there is no reason one could not define further levels of granularity.

The result of this database configuration was that roles could be added to the CAS database as target resources as files could be in the original prototype and users could be given membership right on those roles. This allowed CAS to express rights such as "the user is a member of atlas/admin".

### 4.2.2. Changes to Resource Server for Enforcement

In the CAS model, the services on the resource are responsible for parsing the CAS credentials, determining the rights granted by those credentials and enforcing those rights. Since our prototype changed the CAS credential content from an explicit list of rights to an assertion of role membership, this required changes to these services to understand and enforce role assertions. The CAS prototype [3] we worked with included a CAS-enabled GridFTP server [1] capable of understanding and enforcing normal CAS authorization assertions. We modified this server to understand and handle with our CAS role assertions.

The CAS prototype comes with a set of libraries for allowing services to make authorization queries regarding the rights expressed in the CAS assertions. For example, queries of the form "does the user have the 'read' right on file '/tmp/foo'?" can be asked. These libraries allowed us to query regarding the user's role memberships without modification, we simply asked about the 'membership' right on the role name we were interested in.

We added a simple configuration file on the resource that listed the roles honored and mapped those roles to Unix accounts. The GridFTP server was modified to walk through the list of groups in the configuration file, query the CAS rights to determine if the user had a membership in that role and if so map the user to the account associated with the role for further enforcement.

If no role membership of the user matched any in the configuration file (or the user had no role membership), then the GridFTP application used the user's personal identity to determine the local account to use as normal.

This system served to allow the local administrator to use the configuration file to map roles to account and then use normal local mechanisms (file permissions, quotas, etc.) to manage policy on the role.

Finally as a security precaution we added a check to verify that the CAS server issuing the role assertion had permission to access the account the role mapped to (i.e. the identity of the CAS server was authorized to map to the account via the Globus Toolkit grid-mapfile). In situations where more than one CAS server is accepted by a resource (e.g. it is serving multiple VOs) this would prevent a CAS server from one VO from either maliciously or mistakenly issuing roles assertions for another VO. Alternately we could have made the role to account mappings specific to the CAS server.

### 4.3. Prototype & Testbed Setup

Standard usage of CAS involves three distinct actors (Typically on three different host machines.):
1> The user (ie. Someone who wishes to access resources using rights granted by CAS.) running client software
2> The CAS Server providing the extended X.509 certificate for use by the user
3> The resource server (eg. a Gridftp Server) providing the actual resource

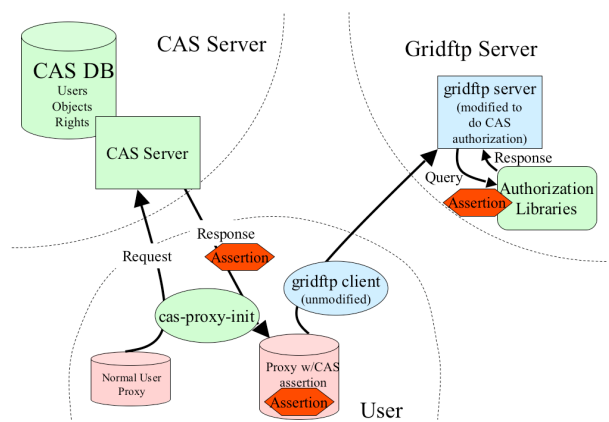

Figure 1. Standard (non-prototype) CAS Architecture in action.

Figure 1 shows the typical interactions when a user accesses a Gridftp server modified to do CAS authorization. The CAS user first gets a standard Grid proxy certificate, then requests credentials from a CAS server. The CAS server replies with credentials based on the rights granted to the user in the CAS database. The user then presents the CAS credentials to the resource server (This is done by running unmodified GSI client applications with a simple wrapper script.) which uses them in making policy decisions about access to resources.

On the user's client machine, a standard CAS session looks something like:

```
# Initialize Grid certificate
% grid-proxy-init
Your identity:
/O=doesciencegrid.org/OU=People/CN=Craig E. Tull 49565
Enter GRID pass phrase for this identity:
Creating proxy... Done
Your proxy is valid until Wed Mar 19 20:30:53 2003
```

```
# Initialize CAS certificate with tag="castag"
% cas-proxy-init castag
# Use CAS tag to connect
% cas-wrap castag gsincftp pdsfgrid3
NcFTP 3.0.3 (April 15, 2001) by Mike Gleason
(ncftp@ncftp.com).
Connecting to 128.55.24.28...
pdsfgrid3.nersc.gov FTP server (GridFTP Server
1.0 -- CAS enabled [GSI patch v0.5] wu-2.6.2(2)
Wed Mar 5 17:42:41 PST 2003) ready.
```

To accomplish the demonstration of CAS as a sub-VO group & role manager, we added new service types to the CAS database and then added new actions to that service type. In our case, we added service type "group" and added action "member". This was trivially done with the standard mechanism for extending the CAS database: An ascii file is created with the appropriate commands, and then loaded into the CAS database.

We then added two new objects of service type "group" to the database. We called the two groups "atlas/admin" and "atlas/data" denoting an administrator group and a data-management group. Users in the CAS database can then be added and removed from the "atlas/data" group by granting and rescinding the "atlas/data" "group" rights to the user in question. This is typically done with the CAS GUI which makes it a trivial process as well.

In our tests, we were operating our modified Gridftp server along side a standard (unmodified) server. This was accomplished by binding the modified server to a different port on the server machine, and specifying that port on the command line when connecting with client software. IP filters and the hosts.allow file were tweaked to allow secure access from the client host.

In a client session, the user can specify the CAS-granted rights that he/she wants associated with a particular CAS tag (The tag is in effect the identifier of which CAS proxy is to be used.) by specifying those rights in an ascii configuration file which is then passed to the cas-proxy-init command via a flag. An example file requesting membership in the atlas/admin group with read access to all appropriate files on the pdsfgrid3 server looks like:

```
# file containing Role privileges
group member atlas/admin wildcard
file read ftp://pdsfgrid3.nersc.gov/* wildcard
```

A user who wants to perform tasks associated with multiple roles or groups would generate multiple CAS proxies, each associated with a tag which associates that proxy and access rights to each subsequent command in the session. A typical user session using the groups capability of CAS might then look like:

```
# Initialize Grid certificate
% grid-proxy-init
Your identity:
/O=doesciencegrid.org/OU=People/CN=George Orwell
1984
Enter GRID pass phrase for this identity:
Creating proxy ... Done
Your proxy is valid until Wed Mar 19 20:30:53
2003
# Initialize 1 CAS certificate for each Role
% cas-proxy-init tull
```

```
% cas-proxy-init -f admin admin
% cas-proxy-init -f data data
# Use Role Tag to connect
% cas-wrap data gsincftp -P 2813 pdsfgrid3
NcFTP 3.0.3 (April 15, 2001) by Mike Gleason
(ncftp@ncftp.com).
Connecting to 128.55.24.28...
pdsfgrid3.nersc.gov FTP server (GridFTP Server
1.0 -- CAS enabled [GSI
patch v0.5] wu-2.6.2(2) Wed Mar 5 17:42:41 PST
2003) ready.
```

Each invocation of a GSI command with the cas-wrap script then specifies the role tag that the user wishes to associate with that command. The cas-proxy-init command without an input file specified simply associates the normal CAS rights for user tull (in this case) with the proxy. A proxy has a limited lifetime, but until it expires, it can be used repeatedly by denoting the appropriate tag as the first argument to cas-wrap.

## 5. SUMMARY, ANALYSIS, & CONCLUSION

### 5.1. Suitability of CAS for this task.

Our tests demonstrate that access to files on a server can be controlled at the sub-VO group or role level by using CAS to create, control, and grant rights associated for that group or role. This allows a single user to adopt multiple rights-sets associated with membership in groups or tenure in a role, and to specify those rights-sets by a simple mechanism (ie. The CAS tag/configuration file mechanism described above.).

In our tests, we chose to interpret the roles and groups as Unix user accounts on the server side. Meaning that multiple individuals come in as the same Unix user for groups, and that the individual associated with a Unix account will change over time for roles. Typical security policies at many computer centers prohibit this kind of "shared account". However, since the X.509 certificate associated with the CAS proxy accessing that accounts contains the distinguished name of the individual accessing the compute resources, auditing of activity on the server host is easily accomplished. This is the main objection that computer centers have to shared accounts.

We conclude that the use of CAS is appropriate for this task. We have not yet tested the use of a modified Gatekeeper for access to CPU resources, but expect that this will be no more complicated than modifying the Gridftp Server.

### 5.2. Integration of Work into Future CAS Releases

In reviewing our modifications, as described in Section 4.2, to the existing CAS prototype we note that the only code changes were in the authorization enforcement code in the resource service (i.e. the GridFTP server). All the changes required to issue role-based assertions was accomplished through simple runtime reconfiguration of the CAS database through the existing administrative interface.

The Globus Project is currently re-implementing the GridFTP server code and plans on allowing for a configurable, modular authorization system that allows for authorization code to be written as dynamic libraries and loaded into the server at runtime. This will allow for code to understand and enforce role-based assertions to be written as a module and placed into a resource service without modifying the resource service, as we were required to do for our prototype.

### 5.3. Standardization of Assertion Format

In modifying CAS to issue an assertion of role membership to a user, it becomes very similar in functionality to the virtual organization membership service (VOMS) [11]. The VOMS system was primarily designed to work with the Globus Toolkit resource management system to allow resources to specify policy based on user's role in a VO. In doing so the VOMS server issues assertions of role membership to users, which they then present to the resource (architecturally identical to the CAS system we used in our prototype).

As with CAS, the VOMS system currently uses a non-standard format for the role-based assertion, requiring custom software on the resource to parse the assertion. By standardizing the format of these assertions we could allow the development of authorization software that could easily understand assertions from either system easily.

Two potential existing standards for such assertion formats already exist, X.509 attribute certificates [5] and security assertion markup language (SAML). While either of these solutions would be appropriate, current research with CAS is investigating SAML due to its current integration with Web services, something that appears to be taking on greater importance in the Grid community.

### 5.4. Use of Unix Group memberships

Our system described in this paper works well for situations where a user only takes on a single role for a given activity. However, it would potentially be the case that a user with authorization to take on different roles may need to do so to accomplish a particular take. For example, if they want to copy a file, readable through their membership in one role, to storage for which they only have write access by membership in another role.

One possible way to allow this is to base role policy not on Unix accounts, but on Unix groups. Since a process (e.g., the GridFTP server) can be a member of multiple Unix groups at a time, this would allow a process to take on rights associated with multiple roles.

### Acknowledgments

This work was supported in part by the Mathematical, Information, and Computational Sciences Division subprogram of the Office of Advanced Scientific Computing Research, Office of Science, U.S. Department of Energy, under contracts W-31-109-Eng-38. Initial funding for CAS was supplied by the Earth Systems Grid project.

This work was supported in part by the Office of Science. Nuclear Physics, U.S. Department of Energy under Contract No. DE-AC03-76SF00098.

The CAS prototype was developed using the pyGlobus from Keith Jackson of LBNL.

The Globus Toolkit is a trademark owned by the University of Chicago.

This is LBNL report number LBNL-52978.